\documentclass[aps,pra,amssymb,amsmath,twocolumn,floatfix]{revtex4-2}
\usepackage{amsmath}
\usepackage{amssymb}
\usepackage{graphicx}
\usepackage{dcolumn}
\usepackage{bm}

\newcommand{\la}{\langle}
\newcommand{\ra}{\rangle}
\newcommand{\w}{\widehat}

\begin{document}

\title{Mobile impurity in a one-dimensional quantum gas: Exact diagonalization in the Bethe Ansatz basis}

\author{Evgeni Burovski}
\affiliation{HSE University, 101000 Moscow, Russia}

\author{Oleksandr Gamayun}
\affiliation{Faculty of Physics, University of Warsaw, ul. Pasteura 5, 02-093 Warsaw, Poland}

\author{Oleg Lychkovskiy}
\affiliation{Skolkovo Institute of Science and Technology,
Bolshoy Boulevard 30, bld. 1, Moscow 121205, Russia}
\affiliation{Department of Mathematical Methods for Quantum Technologies,
Steklov Mathematical Institute of Russian Academy of Sciences
8 Gubkina St., Moscow 119991, Russia}

\begin{abstract}
We consider a mobile impurity particle injected into a one-dimensional quantum gas.
The time evolution of the system strongly depends on whether the mass of the impurity
and the masses of the host particles are equal or not. For equal masses, the
model is Bethe Ansatz solvable, but for unequal masses, the model is no longer
integrable and the Bethe Ansatz technique breaks down. We construct a controllable
numerical method of computing the spectrum of the model with a finite number
of host particles, based on exact diagonalization of the Hamiltonian in the
truncated basis of the Bethe Ansatz states. We illustrate our approach on a
few-body system of 5+1 particles, and trace the evolution of the spectrum
depending on the mass ratio of the impurity and the host particles.
\end{abstract}

\maketitle

\section{\label{sec:level1}Introduction}

Describing the propagation of a foreign particle through a medium---the polaron problem---is a fundamental
physics problem, with a wide variety of applications \cite{Devreese2009, Kuper1963}.
The problem is especially non-trivial for quantum systems at low temperatures and
in reduced dimensions, which enhance quantum fluctuations. Recent advances in
experimental techniques in the field of ultracold gases allow efficient
realization, control and measurements of strongly interacting quantum systems,
including two-component gases with large population imbalance, of which a single
a polaron problem is a limiting case \cite{Palzer2009, Catani2012, Meinert2017, Fukuhara2013}.

The motion of the impurity injected into a one-dimensional (1D) degenerate zero-temperature
quantum gas strongly depends on the ratio of the impurity mass, and the mass of the host particles.
This is already seen from the semiclassical analysis of the kinetic Boltzmann equation,
which predicts that the terminal velocity of the impurity depends on the mass
ratio \cite{GamayunLychkovskiy2014}. For equal masses, the Boltzmann equation
approach fails unless multiple coherent scattering processes are taken into account
\cite{Gamayun2014}.

For equal masses, however, an alternative numerical approach is based on a Bethe Ansatz
solvable model where the host gas consists of non-interacting fermions
and impurity-host interaction is a short-range delta-function repulsion (see Eq.\ \eqref{h0} below).
This model, a special case of a Gaudin-Yang model \cite{Gaudin1967, Yang1967}, is integrable
and the exact spectrum can be computed by solving a system of Bethe Ansatz equations \cite{McGuire1965}.
Explicit expressions for matrix elements for several observables are known \cite{Zotos1993, Mathy2012},
and some physically relevant observables can be computed numerically via summations
over subsets of Bethe Ansatz states \cite{Mathy2012, Burovski2014, Robinson2016, Gamayun2018}.

In this paper, we consider the case where the impurity mass differs from that of the host particles.
While the exact Bethe Ansatz solution is not available, we construct an exact diagonalization
procedure using a suitably truncated basis of the Bethe Ansatz states of an integrable model with equal masses
(so that the off-diagonal matrix elements are related to the mass difference).
Applying our procedure to a mesoscopic system of and impurity and $N=5$ host particles, we study
the evolution of the spectrum with the mass ratio and observe clear signatures of level repulsion
and a redistribution of the spectral weight for $m/M \neq 1$.
Two notes are in order. First, we note that our approach is non-perturbative and does
not rely on either a coupling constant or a mass difference being small. Second,
our numerical procedure is controllable: the truncation error is directly related to the
saturation of the sum rule related to the completeness of the full basis of the Hilbert space.

Our approach is inspired by the so-called Truncated Spectrum Approach (TSA), which is
used to describe perturbations of conformal field theories by relevant operator
(see e.g. Ref.\ \cite{James2018} for a review). In this approach, the subset of eigenstates
that forms a basis for the perturbed Hamiltonian is truncated by their energy values,
which makes the method tailored to capture low-energy physics. The dependence of the
energy cut-off is regulated by the corresponding renormalization group.
In contrast to the TSA approach, we rely on the saturation of a sum rule; this
is similar to the application of
the TSA for the quench description in the Lieb-Liniger model \cite{Caux2012}.
(See {Fig.\ \ref{fig:rho_convergence}} below for a comparison of the ordering strategies.)

It is also instructive to compare our technique to alternative numerical approaches, which can
be extended to the unequal mass problem. One popular approach is to directly simulate
real-time dynamics using time-dependent density-matrix renormalization
group \cite{Knap2014, Massel2013, Peotta2013,Ozaki2015}. Our approach is complementary, as
it is based on a different strategy of approximating the low-lying part of the spectrum.
Directly solving a many-body Schroedinger equation was used in Ref. \cite{Petrov2011} for
studying the response to a parametric modulation of the coupling constant for
both equal and unequal mass systems. This approach is obviously superior for studying
few-particle systems, but quickly becomes impractical for larger number of particles.

The rest of the paper is organized as follows. In Sec.\ \ref{sec:model} we detail the microscopic
model, and discuss our numerical approaches for studying both integrable, equal-mass
(Sec.\ \ref{subsec:eig_int}) and non-integrable, unequal-mass (Sec.\ \ref{subsec:eig_non_int})
cases. Sec.\ \ref{sec:results} presents results of our numerical experiments, and
Sec.\ \ref{sec:concl} concludes the paper.

\section{Model and method}
\label{sec:model}

We consider an impurity of mass $M$ injected into the 1D gas of free spinless fermions
(equivalently, Tonks-Girardeau bosons) at $T=0$.
The host gas contains $N$ identical fermions of mass $m$. The impurity interacts
with the host particles via the short-range delta-function repulsion. Then the
Hamiltonian in the first quantization reads

\begin{equation}
\w{H} = \frac{\w{P}^2}{2M} + \sum_{i=1}^N \frac{\w{p_i}^2}{2m} + \frac{g}{m} \sum_{j=1}^N \delta(X - x_j) \,.
\label{ham_orig}
\end{equation}

Here $X$ is the coordinate of the impurity and $x_j$ is the coordinate of the $j$-th host
particle, $\w{P}$ and $\w{p_j}$ are the momenta operators of the impurity and  $j$-th
host particle, respectively; $g > 0$ is the strength of the delta-function coupling between
host particles and the impurity. In Eq.\ \eqref{ham_orig} and below we use the
units where the Planck's constant $\hbar=1$, and we also set $2m=1$.

For equal masses, $M=m$, the model \eqref{ham_orig} admits the exact solution
via Bethe Ansatz \cite{McGuire1965}. For unequal masses, $m\neq M$, the model is no longer integrable, and the Bethe
Ansatz machinery of \cite{McGuire1965} breaks down.

We identically rewrite \eqref{ham_orig} to separate the integrable part, $\w{H_0}$,
which is nothing but Eq.\ \eqref{ham_orig} with $m=M$:

\begin{gather}
\w{H} = \w{H_0} + \left( \frac{1}{2M} - \frac{1}{2m} \right) \w{P^2} \,,  \label{h1}\\[1em]
\w{H_0} = \frac{\w{P}^2}{2m} + \sum_{i=1}^N \frac{\w{p_i}^2}{2m} + \frac{g}{m} \sum_{j=1}^N \delta(X - x_j) \label{h0}\,,
\end{gather}

Note that the second term in Eq.\ \eqref{h1} is off-diagonal in the basis of eigenstates of Eq.\ \eqref{h0}.


\subsection{Eigenstates of the integrable model}
\label{subsec:eig_int}

In this subsection, we discuss the eigenstates of the integrable model
\eqref{h0}. Following Refs. \cite{McGuire1965, Mathy2012}, we consider $N$ host
particles with periodic boundary conditions on a ring of circumference $L$.
Consider the spectrum of the Hamiltonian \eqref{h0} with a given total momentum,
$Q$. Let $| \psi \ra$ is an eigenstate of $\w{H_0}$ with the energy eigenvalue
$\varepsilon_\psi$: $\w{H_0}|\psi\ra = \varepsilon_\psi |\psi\ra$.
The eigenstates are fully characterized by a set of $N+1$ rapidities,
$z_j$, $j=1, \dots, N+1$. Rapidities are roots of the Bethe equation \cite{McGuire1965},
\begin{equation}
1 / \tan{z_j} = a z_j - c \;,
\label{eq:ctg}
\end{equation}
where $a = 8 / g L$ and the free term $c$ ensures the conservation of the total
momentum of the system,
\begin{equation}
Q = \frac{2}{L}\sum_{j=1}^{N+1} z_j \;.
\label{eq:Q}
\end{equation}

Due to periodicity of the cotangent function, solutions of the Bethe equation
\eqref{eq:ctg} can be written as
\begin{equation}
z_j = \pi n_j - \delta_j\;,
\label{eq:phases}
\end{equation}
where $n_j$ are integers and phase shifts $\delta_j \in [0, \pi)$. Therefore,
an eigenstate of Eq.\ \eqref{h0} with a given value of the total momentum $Q$
is completely defined by specifying an integer partition $\vec{\lambda}$, i.e. a set of
$N+1$ distinct integers, $n_j$,
\begin{equation}
\vec{\lambda} = \{n_j \in \mathbb{Z}\;,\quad j=1,\cdots, N+1\} \;.
\label{eq:partition}
\end{equation}
Without loss of generality, we assume the partitions to be ordered, $n_1 < n_2 < \cdots n_{N+1}$.

The constructive procedure of finding an eigenstate of \eqref{h0} with a given
total momentum $Q$ is then as follows \cite{McGuire1965, Mathy2012}:
(i) fix a partition $\vec{\lambda}$, Eq.\ \eqref{eq:partition}, and (ii) solve
a system of $N+2$ nonlinear equations: $N+1$ Bethe equations \eqref{eq:ctg}
define the phase shifts, $\delta_j$, as functions of $c$; the latter is fixed by
solving Eq.\ \eqref{eq:Q} given the total momentum $Q$.

Given a partition $\vec{\lambda}$, and its corresponding rapidities, $z_j$, the
energy eigenvalue is
\begin{equation}
\varepsilon_\psi = \frac{1}{2m}\frac{4}{L^2}\sum_{j=1}^{N+1} z_j^2 \;,
\label{eq:e0}
\end{equation}
and the wave functions in the coordinate representation can be constructed explicitly
as certain Slater-type determinants \cite{Mathy2012} of size $N+1$.

Given the initial state at $t=0$, $|\Psi_0\ra$, being the product state
of an impurity moving with the momentum $Q$ and the Fermi sea of the host fermions,
the completeness relation for the eigenstates of the Hamiltonian \eqref{h0} reads
\begin{equation}
\sum_{\psi} |\la \Psi_0 | \psi \ra |^2 = 1 \,.
\label{eq:sum_rule}
\end{equation}
Note that the size of the Hilbert space is infinite even for a finite number of
host particles $N$. The challenge is to find a relevant subset of eigenstates,

\begin{equation}
\mathcal{S} = \{ \psi : \sum_{\psi} |\la \Psi_0 | \psi \ra |^2 = 1 -\epsilon \}\;,
\label{eq:subspace}
\end{equation}
which saturates the sum rule to a predefined accuracy $\epsilon$.
The relevance of the various subsets of eigenstates for correlation functions
in the Lieb-Liniger model was recently under active exploration \cite{Panfil2021, Panfil2021_1}.

\subsubsection{Numerical approaches}
\label{sec:numerical}

Several approaches for generating representative subsets $\mathcal{S}$,
Eq.\ \eqref{eq:subspace}, have been devised. Numerical approaches are most
transparently described using the (formal) analogy between Bethe Ansatz
rapidities and spinless (pseudo)fermions. Specifically, rapidities, $z_j$,
can be formally interpreted as quasi-momenta of $N+1$ spinless pseudo-fermions
(see e.g. \cite{Caux2009}). Then, an arbitrary partition
$\vec{\lambda}$ is uniquely specified by listing
pseudo particle-hole pairs relative to the pseudo Fermi sea, which is nothing but
a set of $N+1$ consecutive integers
$$
\vec{\lambda}_0 = \{-(N+1)/2, \cdots, (N-1)/2 \}\;.
$$
For integrable models, it is typical that contributions from states with small
number of pseudo particle-hole pairs dominate \cite{Caux2009} and thus
numerics can be expected to converge with respect to taking into account
states with an increasing number of pseudo particle-hole pairs. In what follows
we call these pseudo particle-hole pairs ``excitations''.

Ref.\ \cite{Mathy2012} performed a brute-force enumeration of up to $10^5$
states with up to three excitations. When states are ordered by the absolute value of
the overlap with the in-state, $\la \Psi_0 | \psi \ra$, the saturation of the sum rule,
Eq.\ \eqref{eq:subspace} with the number of states, $N_s$, is reported to be a power law,
$\epsilon \sim N_s^\alpha$ with some value of the exponent $\alpha$.

Ref.\ \cite{Burovski2014} developed an alternative, \emph{stochastic} enumeration
approach of constructing a subset of states, Eq.\ \eqref{eq:subspace}, where the states
are generated via a Markov process in the space of partitions $\vec{\lambda}$,
based on the Metropolis algorithm \cite{Metropolis} with the transition probabilities
proportional to $|\la \Psi_0 | \psi \ra|^2$. This way, the process automatically finds
states with largest contributions to the sum rule \eqref{eq:sum_rule}. In practice,
the set of updates where we only change two values in a partition $\vec{\lambda}$
are sufficient for a quick convergence of the sum rule, \eqref{eq:subspace}.

The resulting eigenstates of Eq.\ \eqref{h0} can be classified into several families.
In the thermodynamic limit, $N\to\infty$, the sum rule \eqref{eq:sum_rule} is dominated
\cite{Burovski2014}
by the single-parameter family of states $\mathcal{S}_k$, of the states
\begin{equation}
\vec{\lambda}_k = \{ -(N+1)/2, \cdots, (N-1)/2, k \}\;,
\label{eq:dominant_family}
\end{equation}
where the pseudo-hole is fixed at $(N+1)/2$ and the pseudo-particle is located
at $n_{N+1} = k$. Note that the momentum conservation, \eqref{eq:Q}, constraints
$\mathcal{S}_k$---in fact, any single-excitation family with a fixed pseudo-hole
position ---to contain $O(N)$ states.

At any finite $N$, the contribution of other states is non-negligible. This includes
other single-excitation states, the pseudo Fermi sea, $\vec{\lambda}_0$, and multiply
excited states (see the top panel of Fig.\ \ref{fig:overlaps}).

\subsection{Eigenstates of the non-integrable model}
\label{subsec:eig_non_int}

We now turn our attention to the full Hamiltonian, Eq.\ \eqref{ham_orig}.
In the rest of the paper, we use Latin symbols ($|f\ra$ etc) to label its eigenstates and
Greek symbols ($|\psi\ra$ etc) to label eigenstates of the integrable model \eqref{h0}.

Consider a complete basis of the integrable Hamiltonian, $\w{H_0}$, Eq.\ \eqref{h0}.
In this basis, the full Hamiltonian, \eqref{h1}-\eqref{h0} has off-diagonal elements due to
the kinetic energy of the impurity, $\propto\w{P}^2$. Formally the matrix elements of
$\w{H}$ are given by
\begin{equation}
\la \phi | \w{H} | \psi\ra = \varepsilon_\psi \delta_{\phi\psi} +%
 \sum_{\alpha \in \mathcal{S}} \la \phi | \w{P} | \alpha\ra \la \alpha | \w{P} | \psi\ra \left(\frac{1}{2M} - \frac{1}{2m} \right)  \;,
\label{h_matrix}
\end{equation}
where $|\psi\ra$ is an eigenstate of $\w{H}_0$ with the energy eigenvalue $\varepsilon_\psi$, and we
used the completeness relation $\sum_\alpha |\alpha\ra \la \alpha | = 1$.

Eq.\ \eqref{h_matrix} is exact if all eigenstates are included. We, instead, consider
a subspace, Eq.\ \eqref{eq:subspace}, of the full Hilbert space, and diagonalize
Eq.\ \eqref{h_matrix} \emph{in that subspace}.

The resulting eigenvalues of the truncated Hamiltonian \eqref{h_matrix}, $\varepsilon_f$, approximate
the exact eigenvalues of $\eqref{ham_orig}$---and the accuracy of
the approximation is controlled by the value of $\epsilon$, i.e. by the total missing
weight of the discarded eigenstates of the integrable model. We also note that the
total discarded weight is invariant under the diagonalization procedure, which is nothing but
a unitary transformation between two orthonormal bases, $\{ | \psi \ra \}$ and $\{ | f\ra \}$, thus
$\sum_{\psi\in\mathcal{S}} |\la \psi | \Psi_0 \ra |^2 = \sum_{f} \la f | \Psi_0 \ra |^2$.

To summarize our numeric procedure, we (i) enumerate a set of states $\mathcal{S}$
 \eqref{eq:subspace} with a predefined level $\epsilon$ using the Bethe Ansatz machinery
of the previous section; (ii) construct the truncated matrix of the Hamiltonian \eqref{h_matrix},
using the explicit expressions for the matrix elements $\la \phi | \w{P} | \psi\ra$ with
$\phi, \psi \in \mathcal{S}$, which were derived in Ref. \cite{Mathy2012}; and (iii) diagonalize
the resulting matrix numerically.

\section{Results and discussion}
\label{sec:results}

\begin{figure}[htb]
\includegraphics[width=0.99\columnwidth, keepaspectratio=True]{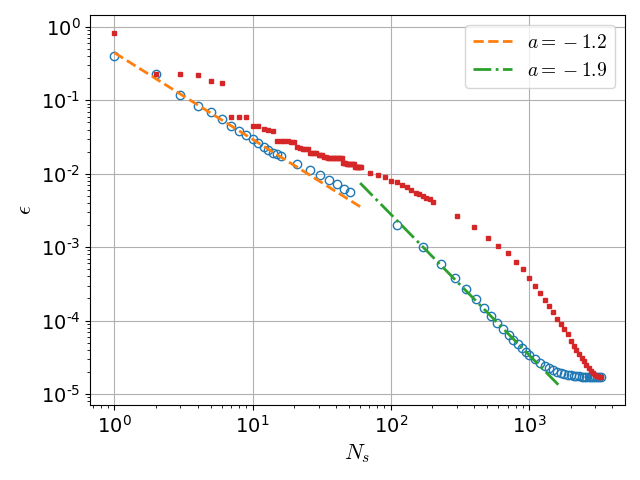} \\%
\caption{\label{fig:rho_convergence} Convergence of the sum rule \eqref{eq:sum_rule} with $N_s$, the number of states
included into the subspace \eqref{eq:subspace}. Open circles show partial sums
of $N_s$ states ordered by the overlap $|\la \psi | \Psi_0\ra|^2$, and filled squares
show partial sums where states are ordered by the energy, Eq.\ \eqref{eq:e0}. Dashed
line illustrates the power-law $\propto N_s^{a}$ with $a=-1.2$ and the dash-dotted line is the
power-law dependence $\propto N_s^{a}$ with $a = -1.9$. See text for discussion.
}
\end{figure}

We first study the convergence of the sum rule, \eqref{eq:sum_rule} , with the
size of the subspace \eqref{eq:subspace}, i.e. the number of states, $N_s$, included into the $\mathcal{S}$.
The dependence of $\epsilon$ on $N_s$ is not unique and depends on which states are
included, not only on their number. Fig. \ref{fig:rho_convergence} shows an illustrative example
where we take $N=5$, $Q = 1.2 k_F$ (here $k_F = \pi N/L$ is the Fermi momentum of $N$ fermions)
and the dimensionless coupling constant $\gamma = g/(N/L) = 4.2$. We generate $3\times 10^3$
eigenstates of Eq.\ \eqref{h0} with up to three excitations using the stochastic enumeration
algorithm of Sec.\ \ref{sec:numerical}, which, collectively, saturate the sum rule \eqref{eq:subspace}
with $\epsilon = 2\times 10^{-5}$.

Fig.\ \ref{fig:rho_convergence} shows partial sums
as a function of $N_s$ for two strategies: where the states are included according
to their energy in the ascending order (red squares), and where the states are included
according their overlap with the $t=0$ state, $|\la \psi | \Psi_0\ra|^2$, in the descending order (open circles).
Clearly, the latter strategy results in much faster convergence of Eq.\ \eqref{eq:subspace}.
In this case, the dependence of $\epsilon$ on $N_s$ has three qualitatively different regimes.
For $10^2 \lesssim N_s \lesssim 10^3$, the convergence can be approximated by a power law
(cf Ref.\ \cite{Mathy2012}), $\epsilon \propto 1/N_s^a$ with $a=1.9$. For smaller
values of $N_s$, the data can be also approximated by a power law, with a smaller exponent, $a\sim 1.2$.
For very large values of $N_s$ (i.e., $\epsilon \lesssim 5\cdot 10^{-5}$),
Fig.\ \ref{fig:rho_convergence} clearly shows convergence which is slower than a power law.
We checked that qualitatively this three-regime behavior with the power-law dependence
with an exponent $a\sim 2$ crossing over to a slower-then-exponential convergence
at large $N_s$ (equivalently, small $\epsilon$), is typical for a wide range of
couplings and larger number of particles.

\begin{figure}[hbt]
\includegraphics[width=0.99\columnwidth, keepaspectratio=True]{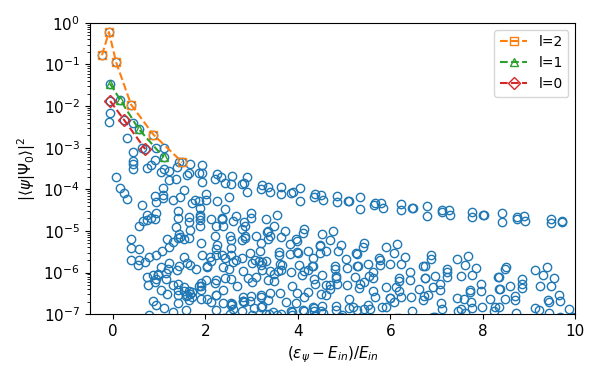} \\%
\includegraphics[width=0.99\columnwidth, keepaspectratio=True]{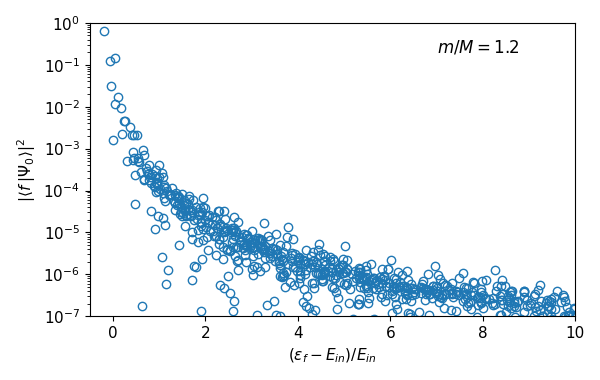} \\%
\caption{\label{fig:overlaps} (top panel) Overlaps, $|\la \psi | \Psi_0\ra|^2$ as a function
of energy, $\varepsilon_\psi$, for eigenstates of the integrable model \eqref{h0}.
Open squares, triangles and diamonds show single-excitation families, similar to
\eqref{eq:dominant_family} with a pseudo-hole fixed at $l=2$, $1$, and $0$, respectively.\\
(bottom panel) Overlaps, $|\la f | \Psi_0\ra|^2$ as a function
of energy, $\varepsilon_f$, for eigenstates of the non-integrable model \eqref{h1}--\eqref{h0},
with $m/M = 1.2$, obtained by diagonalization of Eq.\ \eqref{h_matrix} in the
subspace $\mathcal{S}$ spanned by the states shown in Fig.\ \ref{fig:rho_convergence}.
In both panels we only show states with overlaps larger then $10^{-7}$ and energies
smaller then $11 E_\mathrm{in}$, for clarity.
See text for discussion.
}
\end{figure}

Fig.\ \ref{fig:overlaps}(top) shows individual states of Fig.\ \ref{fig:rho_convergence}.
Specifically, for each state we show its overlap, $|\la \psi | \Psi_0\ra|^2$, vs energy $\varepsilon_\psi$,
relative to the energy of the in-state, $E_\mathrm{in} = \la \Psi_0 | \w{H}_0 |\Psi_0\ra$. We also
indicate special single-excitation families, similar to \eqref{eq:dominant_family},
which are parameterized by the position of the pseudo-particle, $k$, at a fixed location
of the pseudo-hole, $l$. For a finite number of host particles ($N=5$ here), these
families are not separated from the rest of the spectrum---the separation increases
with the number of particles $N$ (cf Fig. 2 of Ref.\ \cite{Burovski2014}).
At finite $N$, doubly excited states have a finite contribution to Eq.\ \eqref{eq:sum_rule},
including a set of states with the overlap $\sim 10^{-4}$, which are visually separated from the
lower-overlap part in Fig.\ \ref{fig:overlaps}(top). The contribution of
three-excitation states to the sum rule, \eqref{eq:sum_rule}, is minor: $\sim 600$ states
present in Figs.\ \ref{fig:rho_convergence}-\ref{fig:overlaps}(top) contribute $\sim 3\times 10^{-5}$
in total.

We now turn our attention to constructing the eigenstates of the full model,
Eq.\ \eqref{h1}--\eqref{h0}. We construct the full Hamiltonian matrix \eqref{h_matrix}
in the subspace $\mathcal{S}$, which is spanned by the states shown in Figs. \ref{fig:rho_convergence}
and \ref{fig:overlaps}, and numerically diagonalize it using standard LAPACK
routines. This procedure generates the set, $\mathcal{S}_f$, of eigenstates
with energy eigenvalues $\varepsilon_f$, and orthonormal eigenvectors, $|f\ra$,
of the Hamiltonian \eqref{ham_orig} in the basis of eigenstates of Eq. \eqref{h0}:
$| f \ra = \sum_{\psi\in\mathcal{S}} | \psi \ra \la \psi| f \ra$.

The bottom panel of Fig.\ \ref{fig:overlaps} shows the result of the diagonalization
procedure in the same subspace as Fig. \ref{fig:rho_convergence}. The overlaps with
the in-state are computed via $\la \Psi_0 | f \ra = \sum_{\psi\in\mathcal{S}} \la \Psi_0 | \psi \ra \la \psi| f \ra$.
Here we take the ratio of masses of host particles and the impurity, $m/M = 1.2$.
The overall structure is similar for other mass ratios.

While the total weight is invariant under the unitary transformation,
$\sum_{\psi\in\mathcal{S}} |\la \psi|\Psi_0 \ra|^2 = \sum_{f\in\mathcal{S}_f} |\la f|\Psi_0 \ra|^2$, it is clear
from Fig.\ \ref{fig:overlaps} that having $M \neq m$ drastically changes the
structure of relative weights of individual states. For a non-integrable model,
\eqref{h1}--\eqref{h0}, the overlaps  with in-state as a function of energy
for the majority of states cluster in a vicinity of a certain master curve,
in a stark contrast to the states of the integrable model, \eqref{h0}.
This situation resembles the typical behavior of thermalization in a generic
isolated quantum system \cite{Rigol2008}. This approach was recently used to describe
observables for the Holstein polaron, which is very close to our case \cite{Jansen2019}.
The eigenstates thermalization hypothesis was clearly demonstrated in the sense
that diagonal matrix elements are functions of the energy only.
We see that in our case for $m\neq M$  similar trend is present even for the
overlaps with the in-state.  We believe that this result might be useful for
the computations of correlation functions.

\begin{figure}[hbt]
\includegraphics[width=0.99\columnwidth, keepaspectratio=True]{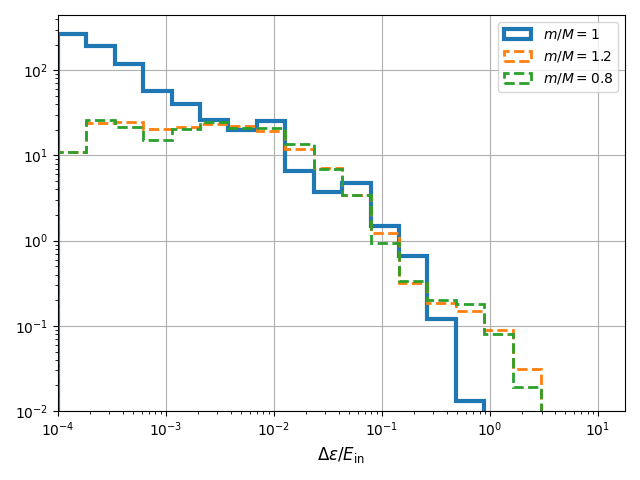} \\%
\caption{\label{fig:NNSD} Nearest neighbor level spacing distributions for
the integrable model, $m=M$ (solid blue line), and non-integrable models with a
heavy impurity, $m/M = 0.8$ and a light impurity, $m/M = 1.2$ (dashed lines).
To set the energy scale, level spacing is given relative to the in-state energy $E_\mathrm{in}$
of the integrable model.
See text for discussion.
}
\end{figure}

To further probe the effect of $M\neq m$, we consider statistics of energy levels. For an
integrable case, $M = m$, the levels are expected to be uncorrelated (the system
is then said to be regular) \cite{Berry1977}, while for $M \neq m$ spectral
statistics is expected to be chaotic. Fig.\ \ref{fig:NNSD} shows the distribution of energy level spacing,
$\Delta\varepsilon = \varepsilon_{j+1} - \varepsilon_{j}$ for both integrable
and non-integrable cases. Here we take the subspace $\mathcal{S}$ of
Fig.\ \ref{fig:rho_convergence}, which contains $3\times 10^3$ states, and diagonalize the
Hamiltonian matrix \eqref{h_matrix} for $m/M = 1.2$ and $m/M = 0.8$. A clear feature of
Fig.\ \ref{fig:NNSD} is a reduction of distribution for
$\Delta\epsilon / E_\mathrm{in} \lesssim 10^{-3}$ for $M\neq m$---which signals
level repulsion, typical for non-integrable systems \cite{Bohigas1984}. We also note that
the shape of the distribution distinguishes between integrable and non-integrable
systems, but is similar for both light or heavy impurity ($M < m$ or $M > m$).

\begin{figure}[htb]
\includegraphics[width=0.99\columnwidth, keepaspectratio=True]{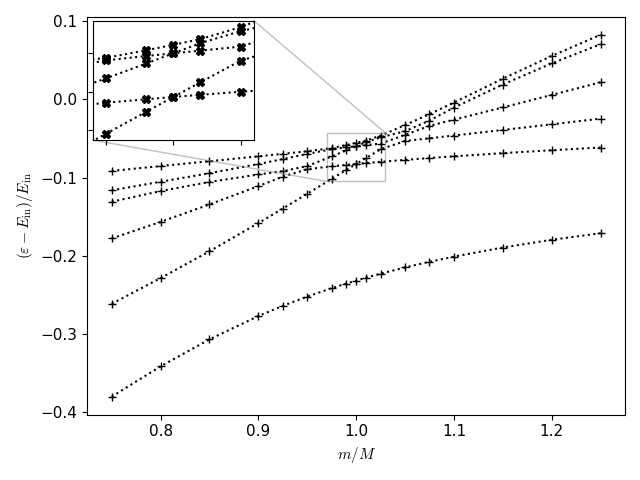} \\%
\caption{\label{fig:ene_mr} Evolution of six lowest-lying energy states with the
mass ratio, $m/M$. To set the energy scale, we scale the energies relative to
the in-state of the integrable model, $E_\mathrm{in}$.
See text for discussion.
}
\end{figure}

Fig.\ \ref{fig:ene_mr} illustrates the dependence of the ground state and five lowest
excited states with the mass ratio, $0.75 \leqslant m/M \leqslant 1.25$. The ground state is non-degenerate and
monotonic with $m/M$; excited states show the Landau-Zener behavior in the vicinity
of $m/M = 1$.

\section{Conclusions and outlook}
\label{sec:concl}

We study a model which captures the physics of an impurity injected into a degenerate 1D
Fermi gas. For equal masses of the impurity and the host particles, the model admits
an exact Bethe Ansatz solution, which we use for studying the non-integrable model
with unequal masses. We construct an exact diagonalization procedure using
truncated bases of the Bethe Ansatz states of the $m=M$ model. Our method is
non-perturbative---we do not rely on either a coupling constant or the mass difference
being small---and controllable: the accuracy of the truncation of the Hilbert space
is controlled by the ``missing weight'' of the discarded configurations, which
is, in turn, controlled by choosing the set of Bethe Ansatz states we include into
the diagonalization.
We illustrate our numerical procedure on a system of $5+1$ particles and compare
spectral properties and redistribution of spectral weights for $m=M$ and $m\neq M$.

Comparing our approach to alternative numerical methods, we note that typically
exact diagonalization and matrix-product state simulations work with
lattice models. (For parabolic traps, a truncated basis of harmonic oscillator
states can also be used \cite{Grass2015}.)
Our approach is, however, fully off-lattice---it can of course
be extended to lattice models which allow constructing a Bethe Ansatz basis to
build the diagonalization procedure on.

For few-body systems off-lattice, the Schroedinger equation can be directly solved
numerically \cite{Petrov2011}. However, this approach is limited to $N < 4$, while
our procedure can be directly extended to mesoscopic systems of a few tens of
particles, as typical in current experiments with mixtures of ultracold quantum
gases \cite{Palzer2009, Catani2012, Meinert2017}. Furthermore, our numerical results
can be useful for benchmarking and extending various approximate analytical and semi-analytical
approaches, see e.g. \cite{Doggen2013,Mehta2014,Dehkharghani2015,Petkovic2015,Das2018,Heitmann2020,Kwasniok2020,Sorout2020,Brauneis2021, Levinsen2021, Vidana2021,Doi2021,Dolgirev2021}.

\textit{Acknowledgments.---} E.B. acknowledges support of the Academic Fund Program
at the NRU Higher School of Economics (HSE) in 2020-2021 (grant No 20-01-025).
O.G. acknowledges support from the Polish National Agency for Academic Exchange (NAWA) through the Grant No. PPN/ULM/2020/1/00247


\end{document}